\documentclass[aps]{revtex4-2}
\usepackage{graphicx} % Required for inserting images
\usepackage{natbib}
\usepackage{color}

\begin{document}
\title{Coherent $\gamma$-ray Generation By Partially Stripped Ion Beams}
\author{Nicola Piovella}
\affiliation{Dipartimento di Fisica "Aldo Pontremoli", Universit\`{a} degli Studi di Milano, Via Celoria 16, I-20133 Milano, Italy \&
INFN Sezione di Milano, Via Celoria 16, I-20133 Milano, Italy}
\author{Gordon R. M. Robb}
\affiliation{Department of Physics, University of Strathclyde, Scottish Universities 
Physics Alliance (SUPA) \& Cockcroft Institute}
\date{\today}
\email{g.r.m.robb@strath.ac.uk}

\begin{abstract}
We describe a scheme for generation of coherent $\gamma$-rays by backscattering intense visible laser light from a beam of partially stripped ions. The scheme
is similar in principle to the proposed Gamma Factory at CERN, with the important difference that the scattering becomes coherent as a result of a collective instability which microbunches the ions.  This instability is analogous to that which occurs in high-gain free electron lasers (FELs). The scheme potentially offers a route to a source of tuneable, coherent $\gamma$-rays, opening up a wide range of possible new applications and opportunities. We find that the parameter requirements for realization of coherent $\gamma$-ray generation regime are considerably more stringent than those proposed for the Gamma Factory, requiring significant increases in the pump laser intensity and possibly the ion beam current.
\end{abstract}

\maketitle

\section{Introduction}

The generation of bright, tunable $\gamma$-ray sources represents a longstanding goal in accelerator and beam physics due to their potential impact across materials science, nuclear physics and particle physics. Recent advances in relativistic ion beams have motivated renewed interest in schemes based interactions between ion beams and lasers, most notably the Gamma Factory project at CERN, which proposes to produce high-energy photons via Doppler upshifting of laser light scattered by partially stripped ions circulating in storage rings \cite{Krasny2015,Budker2020,Budker2021,Budker2026}. Although this approach could potentially produce intense $\gamma$-ray beams, the emitted radiation is fundamentally incoherent, limiting its  coherence properties and spectral brightness.

A natural extension of these concepts is the pursuit of methods for production of a laser-like beam of coherent $\gamma$-rays, analogous to the operation of free-electron lasers (FELs) operating in the x-ray regime \cite{McNeil2007,Pelligrini2016}. In FELs, collective instabilities driven by the interaction between a relativistic electron beam and a radiation field lead to microbunching of the electron beam and exponential amplification of coherent radiation \cite{Saldin1980,BPN,McNeil2007,Pelligrini2016,Emma2010,Altarelli2007,Faatz2010,Shintake2008,Shintake2010,Patterson2010,Kim2022}. An analogous mechanism exists in atomic and ionic systems in the form of the collective atomic recoil laser (CARL), in which the interplay of optical fields and particle motion leads to self-organization via microbunching of the atoms or ions and consequently coherent scattering of the radiation field \cite{Bonifacio1994NIMA,Bonifacio1994PRA,Bonifacio1997,Piovella2001}.

In this work, we investigate the possibility of extending CARL-like dynamics to relativistic beams of partially stripped ions, enabling coherent $\gamma$-ray generation via a collective instability. By considering the interaction of a counter-propagating laser field with a relativistic ion beam, we show that the system can be mapped to a CARL configuration in a suitable co-moving frame, where the scattered and pump fields become quasi-degenerate. In this regime, interference between the fields induces a periodic modulation of the ion density, which in turn enhances coherent backscattering, giving rise to an exponential gain of the $\gamma$-ray field.

We derive a self-consistent theoretical model describing this interaction and analyze the conditions required to achieve the collective, high-gain regime. Particular attention is given to the scaling of the key dimensionless parameter governing the instability and to its dependence on experimentally accessible quantities such as laser intensity, ion beam current, and interaction time. Using parameters relevant to current Gamma Factory proposals, we assess the feasibility of reaching the coherent regime and identify the primary limitations imposed by realistic beam and laser configurations.

Our results indicate that, although coherent $\gamma$-ray generation via partially stripped ion beams is, in principle, achievable, the conditions to be met for its realization are significantly more stringent than those needed for incoherent emission. In particular, significant increases in laser intensity and/or beam current are necessary to trigger the collective instability. These findings will guide future experimental efforts aimed at bridging the gap between incoherent $\gamma$-ray sources and fully coherent laser-like $\gamma$-ray beams.

\section{Coherent $\gamma$-rays from Ion Beams via a Collective Instability}
\subsection{Collective Atomic Recoil Laser (CARL)}
The collective atomic recoil laser (CARL) model \cite{Bonifacio1994NIMA,Bonifacio1994PRA,Bonifacio1997,Piovella2001,vonCube2004,Gisbert2020,Slama2007,Tomczyk2015} describes
collective Rayleigh scattering of light from cold atoms or ions. It was originally developed by assuming a cold thermal atomic gas with zero average velocity and quasi-degenerate pump and scattered optical fields, as would be expected for elastic Rayleigh scattering. It describes a process where interference between pump and scattered fields creates a spatially periodic optical potential which causes bunching of the atoms. These bunched atoms in turn scatter the optical pump beam coherently to amplify the backscattered beam via a collective instability in which many atoms interact collectively via the common optical fields with which they interact. 

The CARL concept was extended to include a relativistic beam of ions in \cite{Bonifacio1997NIMA}. The system behaves very similarly to the case of a zero average-velocity thermal gas of ions when we transform to a frame of reference co-moving or almost co-moving with the ions, as shown schematically in Fig~\ref{fig:ref_frames}. However, in the lab frame the frequencies of the pump field and scattered field are now very different due to the large Doppler shift involved. 

Here we reconsider the relativistic CARL model, as described in \cite{Bonifacio1997NIMA}, and use it to determine the requirements for performing coherent $\gamma$-ray generation by a beam of partially stripped ions like that proposed for the Gamma Factory at CERN.  

\subsection{Model \& Assumptions}
\begin{figure}
    \centering
    \includegraphics[width=0.75\linewidth]{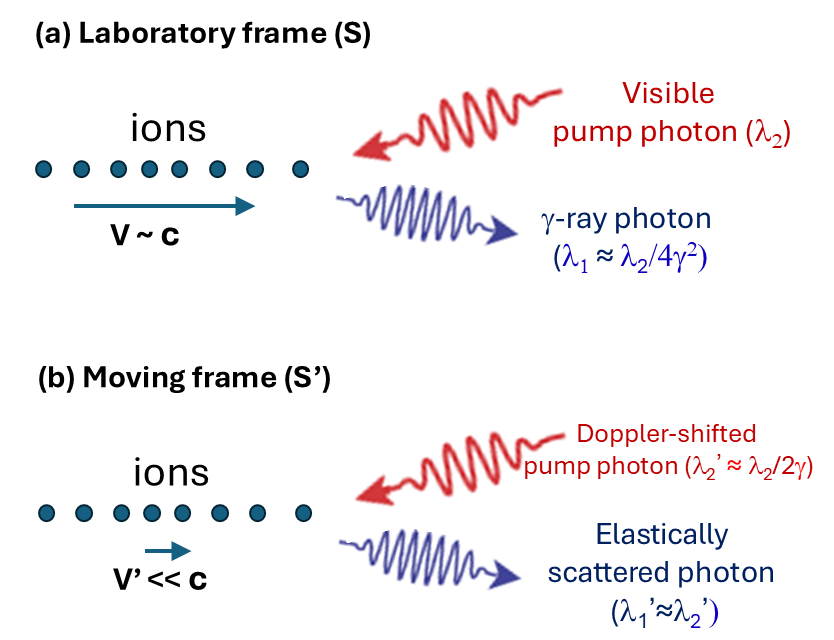}
    \caption{Schematic 1D representation of the $\gamma$-ray generation process in (a) the laboratory frame (S) and (b) in the moving frame (S').}
    \label{fig:ref_frames}
\end{figure}

We consider a mono-energetic beam of two-level ions having a proper transition frequency $\omega_0'$. In the laboratory reference frame, $S$, the ions interact with a strong counter-propagating pump field 
\[
E_2=E_{02}\mathbf{\hat x} \exp[i(k_2z+\omega_2 t)],
\]
where $k_2=\omega_2/c$, and by a weak co-propagating probe field 
\[
E_1=E_{01}\mathbf{\hat x} \exp[i(k_1z-\omega_1 t)] .
\] 
We consider a reference frame $S'$ moving at a velocity $v_r=c\beta_r=c(\omega_1-\omega_2)/(\omega_1+\omega_2)$ where the two frequencies coincide, as shown in Fig.~\ref{fig:ref_frames}. This can be seen using the Lorentz transformations
\begin{eqnarray}
\omega'_{2,1}=\gamma_r\omega_{2,1}(1\pm\beta_r).
\end{eqnarray}
If $\omega'_1=\omega'_2=\omega'$, then $\omega_2(1+\beta_r)=\omega_1(1-\beta_r)$ and $\omega_1=\omega_2(1+\beta_r)/(1-\beta_r)=(1+\beta_r)^2\gamma_r^2\omega_2$. The resonance condition $\omega'=\omega'_0$ implies $\omega_{2,1}=\omega_0'/[\gamma_r(1\pm\beta_r)]$.

\subsection{CARL in the moving frame ($S'$)}
We assume the ion motion to be non-relativistic in the moving frame $S'$. The system is then represented by the following Hamiltonian
\begin{eqnarray}\label{Ham}
    H &=&\sum_{j=1}^N\left\{\frac{\hbar\omega'_0}{2}\sigma_{3j}+\frac{{p'_j}^2}{2m}+\hbar g'\left[{a'_1}^\dagger \sigma_{j}^- e^{-i(k'z'-\omega't')}+
    {a'_2}^\dagger \sigma_{j}^- e^{i(k'z'+\omega't')}+\mathrm{h.c.}\right]\right\}
\end{eqnarray}
where wavenumber $k'=\omega'/c$ and coupling rate $g'=\mu[\omega'/(2\hbar\epsilon_0 V')]^{1/2}$, where $V'$ is the quantization volume of the radiation field in $S'$ and $\mu$ is the electric dipole moment of the transition.  The internal dynamics of the two-level ions are described by the operators
$\sigma_{3j}=|e_j\rangle \langle e_j|- |g_j\rangle \langle g_j|$,~
$\sigma_{j}^+=|e_j\rangle\langle g_j|$ and
$\sigma_{j}^-=|g_j\rangle\langle e_j|$. Furthermore, we also consider the dynamics of external degrees of freedom, where $z'_j$ and $p'_j$ are operators.
The Heisenberg equations are:
\begin{eqnarray}
\frac{dz'_j}{dt} &=& \frac{p'_j}{m},\\
\frac{dp'_j}{dt} &=& 
i\hbar k'g'\left[{a'_1}^\dagger \sigma_{j}^- e^{-i(k'z'-\omega't')}-
    {a'_2}^\dagger \sigma_{j}^- e^{i(k'z'+\omega't')}-\textrm{h.c.}\right],\\
  \frac{d\sigma_{j}^-}{dt}&=& -i\omega'_0\sigma_j^- +ig'
  \sigma_{3j}\left\{ {a'_1}  e^{i(k'z'-\omega't')}+{a'_2} e^{-i(k'z'+\omega't')}\right\},\\
  \frac{d\sigma_{3j}}{dt} &=& 2ig'
  \left[{a'_1}^\dagger\sigma_{j}^- e^{-i(k'z'-\omega't')}+{a'_2}^\dagger \sigma_{j}^- e^{i(k'z'+\omega't')} -\textrm{h.c.}\right],\\
  \frac{da'_1}{dt} &=& -ig'\sum_{j=1}^N \sigma_{j}^-e^{-i(k'z'-\omega't')}.
\end{eqnarray}
where we have assumed that the pump field is sufficiently strong to be represented by a constant real number $a_2$. Furthermore, we  neglect the population of the excited state, assuming $\sigma_{3j}\approx -1$. Introducing the coherence variable $\sigma_j=\sigma_{j}^- e^{i(k'z'+\omega't')}$, the equations become 
\begin{eqnarray}
\frac{dz'_j}{dt}&=& \frac{p'_j}{m},\\
\frac{dp'_j}{dt} &=& 
i\hbar k'g'\left[{a'_1}^\dagger \sigma_{j} e^{-2ik'z'}-
    {a'_2}^\dagger \sigma_{j} -\textrm{h.c.}\right],\\
 \frac{d\sigma_j}{dt} &=& i\left(\Delta'+\frac{k'p'}{m}\right)\sigma_j -
  ig'\left\{ {a'_1}  e^{2ik'z'}+{a'_2}\right\}-\frac{\Gamma'}{2}\sigma_j,\\
 \frac{da'_1}{dt} &=& -ig'\sum_{j=1}^N \sigma_{j}e^{-2ik'z'}.
\end{eqnarray}
where $\Delta'=\omega'-\omega'_0$ is the detuning between the radiation fields and the transition, and we have added the spontaneous emission decay term $-(\Gamma'/2)\sigma_j$, with $\Gamma=\mu^2k'^3/2\pi\epsilon_0\hbar$ as the rate of spontaneous decay. In order to eliminate the internal degrees of freedom we neglect the Doppler shift $k'p'/m$ and $\Gamma'$ with respect to the detuning $\Delta'$ and neglect $a'_1$ with respect to the pump field $a'_2$. For times much larger than $1/\Gamma$ the time derivative can be neglected, $\dot\sigma_j\approx 0$, and we obtain $\sigma_j\approx g'a'_2/\Delta'$ and
\begin{eqnarray}
\frac{d\theta'_j}{dt'}&=& 2k'\frac{p'_j}{m},\label{C1}\\
\frac{dp'_j}{dt'}&=&-
\hbar k'\frac{g'^2a'_2}{\Delta'}\left[{a'} e^{i\theta'_j}+{a'}^\dagger e^{-i\theta'_j}\right],\label{C2}\\
\frac{da'}{dt'} &=& \frac{g'^2a'_2}{\Delta'}\sum_{j=1}^N e^{-i\theta'_j}\label{C3}.
\end{eqnarray}
where we have defined $\theta'=2k' z'$ and $a'=ia'_1$. These are the CARL equations in the moving frame $S'$.

\subsection{Transformation to the laboratory frame (S)}

We now transform Eqs.(\ref{C1})-(\ref{C3}) into the laboratory frame $S$. The phase $\theta'_j$ transforms as
\begin{equation}
\theta'_j=2k'z'=k_1(z_j-ct)+k_2(z_j+ct)=(k_1+k_2)(z_j-c\beta_r t)=\theta_j
\end{equation}
The momentum $p'_j$ transforms as
\begin{equation}
p'_j=\gamma_r(p_j-\beta_r E_j/c)=m\gamma_r\gamma_j(v_j-v_r)
\end{equation}
since $p_j=m\gamma_jv_j$ and $E_j=mc^2\gamma_j$, with $\gamma_j=(1-\beta_j^2)^{-1/2}$ and $\beta_j=v_j/c$. Then, since $dt'=dt/\gamma_r$,
\begin{eqnarray}
\frac{d\theta_j}{dt}&=& 2k_2(1+\beta_r)\gamma_r\gamma_j(v_j-v_r)
\end{eqnarray}
We define $\eta_j=\gamma_r\gamma_j(\beta_j-\beta_r)$, so that
\begin{eqnarray}
\frac{d\theta_j}{dt}&=& 2\omega_2(1+\beta_r)\eta_j
\end{eqnarray}
We observe that in the non relativistic CARL ($\beta_j\ll 1$ and $\beta_r\ll 1$), $\eta_j=(v_j-v_r)/c$, whereas in the ultra-relativistic FEL limit, ($\beta_{j,r}\approx 1-1/2\gamma^2_{j,r}$ and $\gamma_j\sim\gamma_r$), $\eta_j=2(\gamma_j-\gamma_r)/\gamma_r$.

Let now consider the coupling constant $g'=\mu[\omega'/(2\hbar\epsilon_0 V')]^{1/2}$. Since $V'=\gamma_r V$ and $\omega'=\gamma_r\omega_{2,1}(1\pm\beta_r)$, $g'=g_2\sqrt{1+\beta_r}=g_1\sqrt{1-\beta_r}$, where $g_{1,2}=\mu[\omega_{1,2}/(2\hbar\epsilon_0 V)]^{1/2}$.

The detuning transforms as $\Delta'=\gamma_r(1+\beta_r)\omega_2-\omega'_0$. We do not write it in the laboratory values, since it refers to the detuning with respect to the proper transition frequency $\omega'_0$.

Let now consider the transformations for $a'$ and $a'_2$. Since $a^\dagger a=\epsilon_0 E_0^2 V/2\hbar\omega$, and since $E'_{1}=\gamma_r(1-\beta_r)E_1$ and $E'_{2}=\gamma_r(1+\beta_r)E_2$, we have
\begin{eqnarray}
{a'}^\dagger a'&=&a^\dagger a\gamma_r^2(1-\beta_r)\\
{a'}_2^\dagger a'_2&=&a_2^\dagger a_2\gamma_r^2(1+\beta_r)\\
\end{eqnarray}
so that
\begin{eqnarray}
a'&=&\frac{a}{\sqrt{1+\beta_r}}\\
a'_2&=&\frac{a_2}{\sqrt{1-\beta_r}}=a_2\gamma_r\sqrt{1+\beta_r}
\end{eqnarray}
Then, Eqs.(\ref{C1})-(\ref{C3}) transform into:
\begin{eqnarray}
\frac{d\theta_j}{dt}&=& 2\omega_2(1+\beta_r)\eta_j\\
\frac{d\eta_j}{dt} &=& -\frac{\hbar\omega_2}{mc^2}\frac{g_2^2a_2(1+\beta_r)^2\gamma_r}{\Delta'}
\left[{a e^{i\theta_j}}+{a}^\dagger e^{-i\theta_j}\right],\label{C2:S}\\
\frac{da}{dt} &=& \frac{g_2^2a_2(1+\beta_r)^2N}{\Delta'}\langle \exp(-i\theta)\rangle\label{C3:S},
\end{eqnarray}
where $\langle \exp(-i\theta)\rangle=(1/N)\sum_{j=1}^N\exp(-i\theta_j)$.

\subsection{CARL scaling}

We introduce dimensionless time $\tau=2\omega_2(1+\beta_r)\rho t$ and dimensionless momentum $\bar p_j=\eta_j/\rho$,
\begin{eqnarray}
\frac{d\theta_j}{d\tau}&=& \bar p_j\\
\frac{d\bar p_j}{d\tau} &=& -\frac{C_1}{\rho^2}
\left[{a e^{i\theta_j}}+{a}^\dagger e^{-i\theta_j}\right],\label{C2:D}\\
\frac{da}{dt} &=& \frac{C_2}{\rho}\langle \exp(-i\theta)\rangle\label{C3:D},
\end{eqnarray}
where $C_1=[\hbar g_2^2a_2(1+\beta_r)\gamma_r]/(2mc^2 \Delta')$ and $C_2=[g_2^2a_2(1+\beta_r)N]/(2\omega_2\Delta')$. By defining $A=\rho a/C_2=C_1a/\rho^2$ we obtain
\begin{equation}
\rho=(C_1C_2)^{1/3}=\left[\frac{\hbar g_2^4a_2^2(1+\beta_r)^2\gamma_r N}{4mc^2\omega_2\Delta'^2}\right]^{1/3},
\end{equation}
\begin{equation}
A=\sqrt{\frac{C_1}{C_2}}\frac{a}{\sqrt{\rho}}=\sqrt{\frac{\hbar\omega_2\gamma_r}{mc^2 N\rho}}a
\end{equation}
and the equations in the standard universal form:
\begin{eqnarray}
\frac{d\theta_j}{d\tau}&=& \bar p_j\label{C1:U}\\
\frac{d\bar p_j}{d\tau} &=& -
\left[A e^{i\theta_j}+A^\dagger e^{-i\theta_j}\right],\label{C2:U}\\
\frac{dA}{d\tau} &=& \langle \exp(-i\theta)\rangle\label{C3:U}.
\end{eqnarray}

\begin{figure}[ht]
    \centering
    \includegraphics[width=0.75\linewidth]{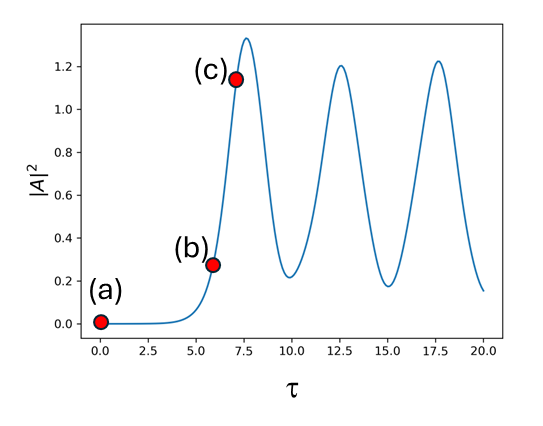}
    \caption{Dimensionless $\gamma$-ray intensity, $|A|^2$, against interaction time $\tau$ as calculated from Eq.(\ref{C1:U}..\ref{C3:U}). The points labeled (a), (b) and (c) relate to the phase-space snapshots shown in Fig.~\ref{fig:phasespace}. 
    }
    \label{fig:intens_vs_t}
\end{figure}

\begin{figure}[ht]
    \centering
    \includegraphics[width=0.75\linewidth]{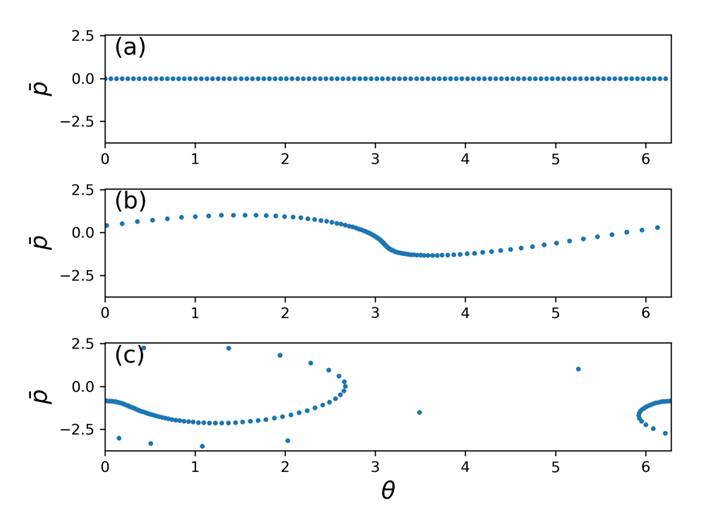}
    \caption{Snapshots of ion phase space showing development of ion bunching. The labels (a), (b) and (c) relate to times at which each snapshot occurs (see Fig.~\ref{fig:intens_vs_t}).
}
    \label{fig:phasespace}
\end{figure}

Figs.~\ref{fig:intens_vs_t} and \ref{fig:phasespace} show the evolution of the ion positions, $\theta_j$, the ion energies, $\bar{p}_j$, and the $\gamma$-ray intensity, $|A|^2$, as calculated from eq.(\ref{C1:U})-(\ref{C3:U}). Fig.~\ref{fig:intens_vs_t} shows that the $\gamma$-ray intensity is exponentially amplified from it's initially small value before saturating and performing quasi-periodic oscillations. The reason for the exponential amplification can be deduced from Fig.~\ref{fig:phasespace}, which shows that the amplification of the $\gamma$-ray intensity occurs simultaneously with the development of a strong density modulation in the ion density. It is the development of this density grating that causes the scattering process to change from a single-particle process involving spontaneous scattering to a collective process involving many ions, interacting cooperatively via the common pump and scattered ($\gamma$-ray) field with which they interact. As the ion density grating develops, it scatters the pump field coherently, amplifying the $\gamma$-ray beam and consequently increasing the size of the ion density modulation, resulting in an instability in which the ion density modulation depth and the $\gamma$-ray beam intensity undergo exponential growth simultaneously. 
This instability is very similar in character to that which underpins high-gain free electron lasers (FELs) \cite{Emma2010,Altarelli2007,Faatz2010,Shintake2008,Shintake2010,Patterson2010,Kim2022} and indeed Eq.~(\ref{C1:U})-(\ref{C3:U}) are of the same form as the equations describing the high-gain FEL \cite{McNeil2007}.

In Figs \ref{fig:intens_vs_t} and \ref{fig:phasespace} it has been assumed that the ion beam is initially completely monoenergetic. The existence of the collective instability relies on the ion beam being sufficiently monoenergetic. The threshold condition for the fractional ion beam energy spread, $\frac{\sigma_E}{E}$ is \cite{BPN} 
\begin{equation}
    \frac{\sigma_E}{E} < \rho .
    \label{eq:gammaspread}
\end{equation}

The parameter $\rho$ and the radiated field $A$ can be written in a different way. We introduce the quantum parameter $q=mc^2\gamma_r/\hbar\omega_1$, where $\omega_1=\omega_2(1+\beta_r)/(1-\beta_r)=\omega_2\gamma_r^2(1+\beta_r)^2$. Then
\begin{equation}
A=a\sqrt{\frac{\hbar\omega_1}{mc^2 \gamma_r(1+\beta_r)^2N\rho}}=\frac{1}{1+\beta_r}\frac{a}{\sqrt{\bar\rho N}}
\end{equation}
where $\bar\rho=q\rho$. So, $\bar\rho(1+\beta_r)^2|A|^2=|a|^2/N$ is the average number of photons emitted per ion and $\bar\rho$ is the quantum efficiency. 

Also, $2g_2a_2=\Omega_2$, where $\Omega_2=\mu E_{02}/\hbar$ is the pump Rabi frequency.
Introducing the plasma frequency $\omega_p=\sqrt{\mu^2\omega_2 n/2\epsilon_0\hbar}$, where $n=N/V$ is the particle density, we can write
\begin{equation}
\rho=\left[\frac{\Omega_2}{\Delta'}\frac{\omega_p}{4\omega_2}(1+\beta_r)\right]^{2/3}\left(\frac{\hbar\omega_2\gamma_r}{mc^2}\right)^{1/3}=
\frac{1}{q^{1/3}}\left[\frac{\Omega_2}{\Delta'}\frac{\omega_p}{4\omega_2}\right]^{2/3}.
\end{equation}
It shows that the parameter $\rho$ scales as $\gamma_r^{-1/3}$, as predicted by Bonifacio and Barletta \cite{Bonifacio1997NIMA}. We also have
\begin{equation}
\bar\rho=q\rho=
\left[\frac{\Omega_2}{\Delta'}\frac{\omega_p}{4\omega_2}q\right]^{2/3}=\left[\frac{\Omega_2}{\Delta'}\frac{\omega_p\gamma_r}{8\omega_{rec}}\right]^{2/3}
\end{equation}
where $\omega_{rec}=\hbar\omega'^2/2mc^2=\hbar\omega_1\omega_2/2mc^2$ is the recoil frequency. Note that $\bar\rho$ reduces to the CARL parameter \cite{Bonifacio1994NIMA} in the non relativistic limit, with $\omega_1=\omega_2$ and $\gamma_r=1$.

We define detuning as the initial value of $\bar p_j(0)$. For a mono-energetic beam,
\begin{equation}
\delta=\bar p(0)=\frac{\gamma_r\gamma_0}{\rho}(\beta_0-\beta_r)=\frac{\gamma_r\gamma_0(1-\beta_0)}{\rho(\omega_1+\omega_2)}
\left[\omega_2\frac{1+\beta_0}{1-\beta_0}-\omega_1\right]=\frac{\gamma_r\gamma_0(1-\beta_0)(1+\beta_r)}{2\rho\omega_1}
\left[\omega_2\frac{1+\beta_0}{1-\beta_0}-\omega_1\right]
\end{equation}
where we used the expression $\beta_r=(\omega_1-\omega_2)/(\omega_1+\omega_2)$. If $\beta_r\approx \beta_0$, then
\begin{equation}
\delta\approx\frac{1}{2\rho\omega_1}
\left[\omega_2\frac{1+\beta_0}{1-\beta_0}-\omega_1\right]
\end{equation}
The resonance condition $\delta=0$ implies $\omega_1=\omega_2(1+\beta_0)/(1-\beta_0)$. In this case, the intensity increases exponentially as $\exp(\sqrt{3}\tau)$. Using $t=z/c\beta_0$, the growth is as $\exp(z/l_g)$ where the gain length is
\begin{equation}
l_g=\frac{c\beta_0}{2\sqrt{3}\omega_2(1+\beta_r)\rho}=\frac{c\beta_0\gamma_r}{2\sqrt{3}\omega'\rho}=\frac{c\beta_0\gamma_r}{2\sqrt{3}\omega'_0\rho}
\end{equation}
where we assumed $\omega'=\omega'_0$.

The emitted power is
\begin{equation}
P_{out}=\rho P_{beam} |A|^2\frac{(1+\beta_r)^2}{\beta_r}=\bar\rho I(A)E_{photon}(eV) |A|^2\frac{(1+\beta_r)^2}{\beta_r}
\label{eqn:Pout}
\end{equation}
where $P_{beam}(W)=I(A)E_{beam}(eV)$ is the beam power and $E_{photon}(eV)=\hbar\omega_1$ is the photon energy. Notice that $\rho(1+\beta_r)^2$ is the maximum energy efficiency, whereas $\bar\rho(1+\beta_r)^2$ is the maximum number of photons emitted per ion.

The CARL model is valid in $S'$ for (a) $\Delta'\gg\Gamma'$ and (b)  $a'_2\gg a'_1$. This condition in the laboratory frame brings to
\begin{equation}
P_2\gg \left(\frac{1-\beta_r}{1+\beta_r}\right)^2 P_{out}.
\end{equation}
About the particle beam, we take the radius of the beam as $r_b=(\beta^*\epsilon_n/\gamma_r)^{1/2}$, where $\beta^*$ is the beta function and $\epsilon_n$ is the normalized emittance. We assume $\beta^*=l_g$ and the beam current $i_{beam}=ec\beta_r n\pi r_b^2$. Hence,
\[
n=\frac{i_{beam}\gamma_r}{ec\beta_r\pi\epsilon_n\beta^*}=\frac{i_{beam}\gamma_r}{ec\beta_r\pi\epsilon_n}\frac{2\sqrt{3}\omega'\rho}{c\beta_r\gamma_r}
=\frac{2\sqrt{3}i_{beam}\omega'}{ec^2\beta_r^2\pi\epsilon_n}\rho
\]
By inserting it in the expression of $\rho^3$:
\[
\rho^3=\frac{1}{q}\left[\frac{\Omega_2}{\Delta'}\frac{\omega_p}{4\omega_2}\right]^{2}=
\frac{1}{q}\left[\frac{\Omega_2}{4\Delta'\omega_2}\right]^{2}\frac{\mu^2\omega_2}{2\hbar\epsilon_0}n
=\frac{1}{q}\left[\frac{\Omega_2}{\Delta'}\right]^{2}\frac{\mu^2}{\hbar\epsilon_0}
\frac{\sqrt{3}i_{beam}\omega'}{16\omega_2ec^2\beta_r^2\pi\epsilon_n}\rho
\]
and
\begin{equation}
\rho=\frac{\mu}{4c\beta_r}\left[\frac{\Omega_2}{\Delta'}\right]
\sqrt{\frac{\sqrt{3}i_{beam}\omega'_0}{q\omega_2e\pi\hbar\epsilon_0\epsilon_n}}\label{rho:self}
\end{equation}

\section{Requirements for coherent $\gamma$-ray generation}

The proof of principle experiment at CERN will use a beam of ions $^{208}Pb^{79+}$, with a transition $2s\rightarrow 2p_{1/2}$ at $230.81$ eV and a lifetime of the excited state $\tau=76.6$ ps. The relativistic Lorentz factor is $\gamma=96.3$ and the ion mass $m=193.687\, GeV/c^2$. The number of ions per bunch is $N=10^8$, the length of the bunch $\tau_{beam}=213$ ps, the normalized emittance $\epsilon_n=1.5$ mm mrad and the energy spread $\sigma_E/E=2\times 10^{-4}$.

The resonance frequency is $\omega'_0=3.5066\times 10^{17} \mbox{s}^{-1}$, corresponding to a wavelength $\lambda'_0=5.375$ nm. The pump frequency is $\omega_2=\omega'_0/2\gamma=1.82\times 10^{15} \mbox{s}^{-1}$, with wavelength $\lambda_2=2\gamma\lambda'_0=1035.22$ nm. The output frequency is $\omega_1=2\gamma\omega'_0=6.753\times 10^{19} \mbox{s}^{-1}$ (i.e. $\hbar\omega_1=44.454$ keV). The decay rate is $\Gamma'=1/\tau=1.305\times 10^{10} \mbox{s}^{-1}$.
A detuning of $\delta_a=\Delta'/\Gamma'=10^3$ corresponds to a relative frequency shift of about $10^{-4}$, so it can be neglected in the remaining expressions.
Assuming $\Gamma'=\mu^2{\omega'_0}^3/3\pi\hbar\epsilon_0 c^3$, we evaluate the electric dipole as $\mu=2.68\times 10^{-31}$ C m. The other derived parameters are the following:
\begin{itemize}
\item $q=mc^2\gamma/\hbar\omega_1=4.2\times 10^8$;
\item $\omega_{rec}=\hbar\omega_1\omega_2/2mc^2=(\hbar\omega'_0)^2/2mc^2\hbar=2.08\times 10^{8} \mbox{s}^{-1}$ ;
\item $\omega_p=\sqrt{\mu^2\omega_2 n/2\epsilon_0\hbar}=(\lambda'_0/4\pi)\sqrt{3\pi c n/\gamma\tau}=0.265\sqrt{n[m^{-3}]} \mbox{s}^{-1}$;
\item $\rho=q^{-1/3}(\Omega_2/\Delta')^{2/3}(\omega_p/4\omega_2)^{2/3}=0.8\times 10^{-13}(\Omega_2/\Delta')^{2/3}(n[m^{-3}])^{1/3}$;
\item $\bar\rho=q\rho=3.32\times 10^{-5}(\Omega_2/\Delta')^{2/3}(n[m^{-3}])^{1/3}$;
\item $\Omega_2=\mu E_2/\hbar=2.54\times 10^3 \times E_2[V/m] \mbox{s}^{-1}$;
\item $\Omega_2/\Delta'=1.94\times 10^{-7}\times (E_2[V/m]/\delta_a)$;
\item $\rho=0.27\times 10^{-17}\times(E_2[V/m]/\delta_a)^{2/3}\times(n[m^{-3}])^{1/3}$;
\item $l_g=c  \gamma/2\sqrt{3}\omega'_0 \rho=2.4\times 10^{-8}/\rho$ m;
\item $t_g=\gamma/2\sqrt{3}\omega'_0\rho=8.0\times 10^{-17}/\rho$ s.
\end{itemize}
The laser delivers a pulse of energy $5$mJ for $\tau_L=2.8$ps, i.e. a power $P_L=1.8$GW. Since $E_2=\sqrt{2I_L/\epsilon_0 c}=27\times I_L^{1/2}[W/m^2]$ and the laser beam has a transverse size $\sigma_T=0.65$ mm, $I_L=P_L/\pi\sigma_T^2\sim 10^{15} W/m^2$ and $E_2\sim 10^9 V/m$. If $n\sim 10^{18} m^{-3}$, we have $\rho\approx 0.3\times 10^{-5}\times\delta_a^{-2/3}$. For $\delta_a\sim 10^{3}$ we obtain $\rho\sim 10^{-8}$, $\bar\rho\sim 4$ and a gain length $l_g\sim 2.4$m.

We can obtain a better estimate of $\rho$ using the self-consistent expression (\ref{rho:self}). Assuming a beam current $i_{beam}=eN/\tau_{beam}=0.075$ A, we obtain, for the parameters above, $\rho=0.4\times 10^{-6}/\delta_a$.
%The typical value of $\rho$ is $10^{-8}$ and the gain time, $t_g \sim 10ns$ which corresponds to an interaction time of $t_{\rm int} \approx l_g/c \approx 1 ns$, which is significantly longer than the pump duration considered. 

Based on the parameters considered above, the scattering process will be spontaneous in nature and will not reach the collective regime of exponential growth of the $\gamma$-ray beam intensity and ion bunching due to the condition for a monoenergetic ion beam, Eq.~(\ref{eq:gammaspread}), not being satisfied, as typical values of $\rho \approx 10^{-8}$ and $\frac{\sigma_E}{E} \approx 10^{-4}$. 
From the definition of $\rho$ in Eq.(\ref{rho:self}), it can be seen that
\begin{equation}
\rho \propto \left( I_L i_{\rm beam} \right)^{1/2} 
\label{eq:rhoscale}    
\end{equation}
so in order to satisfy Eq.~(\ref{eq:gammaspread}) for the ion-beam proposed for the Gamma Factory, the laser intensity $I_L$ would require to be increased by a factor of $\sim 10^8$ to $I_L \sim 10^{23} W m^{-2} = 10^{19} W cm^{-2}$. The requirements on the pump intensity could be relaxed somewhat in principle if the ion beam current was increased by increasing the number of ions in each bunch or reducing the bunch duration.
In addition, the pump laser duration should be sufficient to reach the collective regime of scattering with exponential growth of the $\gamma$-ray beam intensity and strong ion microbunching (approximately points (b) and (c) in Figs~\ref{fig:intens_vs_t} and \ref{fig:phasespace}), which requires $\tau > 1$ i.e. $t_{\rm int} > t_g$. For $\rho \approx 10^{-4}$, this condition corresponds to $t_{\rm int} > 1 \rm{ps}$.  
%it can be seen from Fig.~\ref{fig:intens_vs_t} that it will be necessary to significantly increase the value of the dimensionless interaction time, $\tau$, so that $\tau \gg 1$. From the definition of $\tau$, it can be seen that 
%\[
%\tau \propto \left( I_L i_{\rm beam} %\right)^{1/2} \; t_{\rm int}
%\]
%so attainment of collective $\gamma$-ray generation would require significant increases of one or more of: pump laser duration (interaction time, $t_{\rm int}$), pump laser intensity ($I_L$) and ion beam current ($i_{\rm beam}$) to the extent that the value of $\left( I_L i_{\rm beam} \right)^{1/3} t_{\rm int}$ is increased by at least an order of magnitude relative to the value corresponding to the proposed parameters for the Gamma Factory. {\color{blue} If the interaction time $t_{\rm int}$ alone could be increased to reach the maximum $\gamma$-ray intensity ($\tau \approx 7.5$ in Fig.~\ref{fig:intens_vs_t}), then for the parameters considered above, the $\gamma$-ray power from Eq.~(\ref{eqn:Pout}) would be $P_{\rm out} \approx 80 \rm{kW}$, corresponding to a $\gamma$-ray photon flux $\sim 10^{19} \rm{s}^{-1}$. As Eq.~(\ref{eqn:Pout}) shows that $P_{\rm out} \propto \rho \propto \left(P_L i_{\rm beam} \right)^{1/2}$, increasing the pump laser power or ion beam current will also increase the $\gamma$-ray photon flux.} 

\section{Conclusions}
We have presented a theoretical study of a scheme for generation of coherent $\gamma$-rays by scattering an intense laser from a beam of partially stripped ions traveling at relativistic velocities. The scheme is based on the proposed Gamma Factory at CERN, which aims to generate incoherent $\gamma$-rays using a similar approach. The essential difference between the Gamma Factory and the scheme described here for coherent $\gamma$-ray generation is that coherent $\gamma$-ray generation relies on 
a collective instability involving back-action of the spontaneously generated $\gamma$-rays to produce bunching of the ions, resulting in exponential amplification of the $\gamma$-rays and the ion density modulation. The requirements for realization of this collective instability and the coherent $\gamma$-ray generation regime are, unsurprisingly, more stringent than those required for the Gamma Factory proposal for incoherent $\gamma$-rays generated spontaneously. In order to attain a regime of coherent $\gamma$-ray generation using the partially stripped ion beams proposed at CERN will require significant increases in  the pump laser intensity and possibly the ion beam current, relative to the values currently proposed for the Gamma Factory.

\bibliography{relativisticCARLNotes}

@article{BPN,
title = {Collective instabilities and high-gain regime in a free electron laser},
journal = {Optics Communications},
volume = {50},
number = {6},
pages = {373-378},
year = {1984},
issn = {0030-4018},
doi = {https://doi.org/10.1016/0030-4018(84)90105-6},
url = {https://www.sciencedirect.com/science/article/pii/0030401884901056},
author = {R. Bonifacio and C. Pellegrini and L.M. Narducci}
}

@article{Altarelli2007,
  title={The European X-ray free-electron laser},
  journal={Technical design report},
  author={Altarelli, Massimo and Brinkmann, Reinhard and Chergui, Majed},
  year={2007}
}

@article{Bonifacio1994NIMA,
title = {Collective atomic recoil laser (CARL) optical gain without inversion by collective atomic recoil and self-bunching of two-level atoms},
journal = {Nuclear Instruments and Methods in Physics Research Section A: Accelerators, Spectrometers, Detectors and Associated Equipment},
volume = {341},
number = {1},
pages = {360-362},
year = {1994},
issn = {0168-9002},
doi = {https://doi.org/10.1016/0168-9002(94)90382-4},
url = {https://www.sciencedirect.com/science/article/pii/0168900294903824},
author = {R. Bonifacio and L. {De Salvo}}
}

@article{Bonifacio1994PRA,
  title = {Exponential gain and self-bunching in a collective atomic recoil laser},
  author = {Bonifacio, R. and De Salvo, L. and Narducci, L. M. and D'Angelo, E. J.},
  journal = {Phys. Rev. A},
  volume = {50},
  issue = {2},
  pages = {1716--1724},
  numpages = {0},
  year = {1994},
  month = {Aug},
  publisher = {American Physical Society},
  doi = {10.1103/PhysRevA.50.1716},
  url = {https://link.aps.org/doi/10.1103/PhysRevA.50.1716}
}

@article{Bonifacio1997,
  title = {Propagation, cavity, and Doppler-broadening effects in the collective atomic recoil laser},
  author = {Bonifacio, R. and Robb, G. R. M. and McNeil, B. W. J.},
  journal = {Phys. Rev. A},
  volume = {56},
  issue = {1},
  pages = {912--924},
  numpages = {0},
  year = {1997},
  month = {Jul},
  publisher = {American Physical Society},
  doi = {10.1103/PhysRevA.56.912},
  url = {https://link.aps.org/doi/10.1103/PhysRevA.56.912}
}

@article{BONIFACIO1997NIMA,
title = {Relativistic theory of the collective atomic recoil laser},
journal = {Nuclear Instruments and Methods in Physics Research Section A: Accelerators, Spectrometers, Detectors and Associated Equipment},
volume = {384},
number = {2},
pages = {337-341},
year = {1997},
issn = {0168-9002},
doi = {https://doi.org/10.1016/S0168-9002(96)00849-2},
url = {https://www.sciencedirect.com/science/article/pii/S0168900296008492},
author = {R. Bonifacio and L. {De Salvo} and W.A. Barletta}

}

@article{Budker2020,
    author = "Budker, Dmitry and Crespo L{\'o}pez-Urrutia, Jos{\'e} R. and Derevianko, Andrei and Flambaum, Victor V. and Krasny, Mieczyslaw Witold and Petrenko, Alexey and Pustelny, Szymon and Surzhykov, Andrey and Yerokhin, Vladimir A. and Zolotorev, Max",
    title = "{Atomic Physics Studies at the Gamma Factory at CERN}",
    eprint = "2003.03855",
    archivePrefix = "arXiv",
    primaryClass = "physics.atom-ph",
    doi = "10.1002/andp.202000204",
    journal = "Annalen Phys.",
    volume = "532",
    number = "8",
    pages = "2000204",
    year = "2020"
}

@unpublished{Budker2021,
    author = "Budker, Dmitry and others",
    title = "{Expanding Nuclear Physics Horizons with the Gamma Factory}",
    eprint = "2106.06584",
    archivePrefix = "arXiv",
    primaryClass = "nucl-ex",
    doi = "10.1002/andp.202100284",
    month = "6",
    year = "2021"
}

@article{Budker2026,
author = {Budker, Dmitry and Krasny, Mieczyslaw Witold and Dutheil, Yann},
title = {Gamma Factory: The light into the future},
journal = {International Journal of Modern Physics A},
volume = {0},
number = {0},
pages = {2649003},
year = {0},
doi = {10.1142/S0217751X26490035},

URL = {https://doi.org/10.1142/S0217751X26490035
},
eprint = {       https://doi.org/10.1142/S0217751X26490035
}

}

@article{Emma2010,
   author ="Emma, Paul and others",
   title="{First lasing and operation of an ångstrom-wavelength free-electron laser}",
   year={2010},
   journal={Nature Photonics},
   pages ={641-647},
   volume = {4},
   doi={10.1038/nphoton.2010.176}
}

@article{Faatz2010,
  title={FLASH II: a seeded future at FLASH},
  author={Faatz, B and Baboi, N and Ayvazyan, V and Balandin, V and Decking, W and Duesterer, S and Eckoldt, HJ and Feldhaus, J and Golubeva, N and Koerfer, M and others},
  journal={Proceedings of IPAC, Kyoto, Japan},
  year={2010}
}

@article{Gisbert2020,
  title={Multimode collective atomic recoil lasing in free space},
  author={Gisbert, Angel T and Piovella, Nicola},
  journal={Atoms},
  volume={8},
  number={4},
  pages={93},
  year={2020},
  publisher={MDPI}
}

@article{Kim2022,
   author = {Kim, Changbum and others},
   year = {2022},
   date= {2022/06/01},
   title ={Review of technical achievements in PAL-XFEL},
   journal = {AAPPS Bulletin},
   page = {15 - 32},
   doi= {10.1007/s43673-022-00045-4}
}

@unpublished{Krasny2015,
    author = "Krasny, Mieczyslaw Witold",
    title = "{The Gamma Factory proposal for CERN}",
    eprint = "1511.07794",
    archivePrefix = "arXiv",
    primaryClass = "hep-ex",
    month = "11",
    year = "2015"
}

@article{McNeil2007,
   author = {McNeil, Brian W. J. and Thompson, Neil R.},
   year={2010},
   date= {2010/12/01},
   title = {X-ray free-electron lasers},
   journal = {Nature Photonics},
   pages={814 - 821},
   volume = {4},
   doi= {10.1038/nphoton.2010.239}
}

@article{Patterson2010,
doi = {10.1088/1367-2630/12/3/035012},
url = {https://doi.org/10.1088/1367-2630/12/3/035012},
year = {2010},
month = {mar},
publisher = {},
volume = {12},
number = {3},
pages = {035012},
author = {Patterson, B D and others},
title = {Coherent science at the SwissFEL x-ray laser},
journal = {New Journal of Physics}
}

@article{Pelligrini2016,
  title = {The physics of x-ray free-electron lasers},
  author = {Pellegrini, C. and Marinelli, A. and Reiche, S.},
  journal = {Rev. Mod. Phys.},
  volume = {88},
  issue = {1},
  pages = {015006},
  numpages = {55},
  year = {2016},
  month = {Mar},
  publisher = {American Physical Society},
  doi = {10.1103/RevModPhys.88.015006},
  url = {https://link.aps.org/doi/10.1103/RevModPhys.88.015006}
}

@article{Piovella2001,
title = {Superradiant light scattering and grating formation in cold atomic vapours},
journal = {Optics Communications},
volume = {187},
number = {1},
pages = {165-170},
year = {2001},
issn = {0030-4018},
doi = {https://doi.org/10.1016/S0030-4018(00)01106-8},
url = {https://www.sciencedirect.com/science/article/pii/S0030401800011068},
author = {N. Piovella and R. Bonifacio and B.W.J. McNeil and G.R.M. Robb}
}

@article{Saldin1980,
  title={Generating of coherent radiation by a relativistic electron beam in an ondulator},
  author={Kondratenko, AM and Saldin, EL},
  journal={Part. Accel.},
  volume={10},
  pages={207--216},
  year={1980}
}

@inproceedings{Shintake2010,
  title={Status report on Japanese XFEL construction project at SPring-8},
  author={Shintake, Tsumoru and others},
  booktitle={Proceedings of the 2010 International Particle Accelerator Conference},
  year={2010},
  organization={IEEE}
}

@article{Shintake2008,
  title={A compact free-electron laser for generating coherent radiation in the extreme ultraviolet region},
  author={Shintake, Tsumoru and Tanaka, Hitoshi and Hara, Toru and Tanaka, Takashi and Togawa, Kazuaki and Yabashi, Makina and Otake, Yuji and Asano, Yoshihiro and Bizen, Teruhiko and Fukui, Toru and others},
  journal={Nature Photonics},
  volume={2},
  number={9},
  pages={555--559},
  year={2008},
  publisher={Nature Publishing Group UK London}
}

@article{Slama2007,
  title = {Superradiant Rayleigh Scattering and Collective Atomic Recoil Lasing in a Ring Cavity},
  author = {Slama, S. and Bux, S. and Krenz, G. and Zimmermann, C. and Courteille, Ph. W.},
  journal = {Phys. Rev. Lett.},
  volume = {98},
  issue = {5},
  pages = {053603},
  numpages = {4},
  year = {2007},
  month = {Feb},
  publisher = {American Physical Society},
  doi = {10.1103/PhysRevLett.98.053603},
  url = {https://link.aps.org/doi/10.1103/PhysRevLett.98.053603}
}

@article{Tomczyk2015,
  title = {Stability diagram of the collective atomic recoil laser with thermal atoms},
  author = {Tomczyk, H. and Schmidt, D. and Georges, C. and Slama, S. and Zimmermann, C.},
  journal = {Phys. Rev. A},
  volume = {91},
  issue = {6},
  pages = {063837},
  numpages = {5},
  year = {2015},
  month = {Jun},
  publisher = {American Physical Society},
  doi = {10.1103/PhysRevA.91.063837},
  url = {https://link.aps.org/doi/10.1103/PhysRevA.91.063837}
}

@article{vonCube2004,
  title = {Self-Synchronization and Dissipation-Induced Threshold in Collective Atomic Recoil Lasing},
  author = {von Cube, C. and Slama, S. and Kruse, D. and Zimmermann, C. and Courteille, Ph. W. and     Robb, G. R. M. and Piovella, N. and Bonifacio, R.},
  journal = {Phys. Rev. Lett.},
  volume = {93},
  issue = {8},
  pages = {083601},
  numpages = {4},
  year = {2004},
  month = {Aug},
  publisher = {American Physical Society},
  doi = {10.1103/PhysRevLett.93.083601},
  url = {https://link.aps.org/doi/10.1103/PhysRevLett.93.083601}
}

\end{document}